\documentstyle[aps]{revtex}

\newcommand{\pint}{\int d^3p\:}
\newcommand{\qint}{\int d^3q\:}
\renewcommand{\vec}[1]{\mbox{\boldmath$#1$}}

\begin{document}
\title{Gauge Theory for a Doped Antiferromagnet in a Rotating
Reference-Frame}
\author{C. K\"ubert and A. Muramatsu}
\address{Institut f\"ur Physik, Universit\"at Augsburg \\
Memmingerstr.\ 6, 86135 Augsburg, Germany}
\date{May 5, 1995}
\maketitle

\begin{abstract}
We study a doped antiferromagnet (AF) using a rotating reference-frame.
Whereas in the laboratory reference-frame with a globally fixed
spin-quantization axis (SQA) the long-wavelength, low-energy physics is given
by the O(3) non-linear $\sigma$-model with current-current interactions
between the fermionic degrees of freedom and the order-parameter field for the
spin-background, an alternative description in form of an U(1) gauge theory
can be derived by choosing the SQA defined by the local direction of the
order-parameter field via a SU(2) rotation of the fermionic spinor.
Within a large-$N$ expansion of this U(1) gauge theory we obtain the phase
diagram for the doped AF and identify the relevant terms due to doping that
lead to a quantum phase transition at $T=0$ from the antiferromagnetically
ordered N\'eel phase to the quantum-disordered (QD) spin-liquid phase.
Furthermore, we calculate the propagator of the corresponding U(1) gauge field,
which mediates a long-range transverse interaction between the bosonic
and fermionic fields. It is found that the strength of the propagator is
proportional to the gap of the spin-excitations. Therefore, we expect as a
consequence of this long-range interaction the formation of bound states
when the spin-gap opens, i.e.\ in the QD spin-liquid phase. The possible
bound states are spin-waves with a (spin-) gap in the excitation spectrum,
spinless fermions and pairs of fermions. Thus, an alternative
picture for charge-spin separation emerges, with composite charge-separated
excitations. Moreover, the present treatment shows an intimate connection
between the opening of the spin-gap and charge-spin separation as well as
pairing.
\\[05mm]
\end{abstract}

\section{Introduction}
\label{Introduction}
Recently much attention has been focused on the problem of understanding the
magnetic as well as transport properties of high-temperature superconductors
(HTSC) as a function of doping. The undoped parent compounds like
La$_2$CuO$_4$ are antiferromagnetic insulators with $S=1/2$ local moments on
the copper-sites \cite{endoh}. Upon doping charge-carriers into the CuO$_2$
planes, the long-range magnetic order is destroyed and these materials become
superconducting at low temperatures, while in the normal state
unconventional properties are observed \cite{unconventional}. The particular
interesting  question arising here is whether these anomalous features are
directly related to the proximity to the magnetic quantum phase-transition or
not.
\par
Since in the relevant low-doping regime the spin-spin
correlation length $\xi$ is large \cite{large xi}, it is appropriate
to address this question from a field-theoretic point of view.
Therefore we will study this problem within an effective long-wavelength,
low-frequency theory, which can be derived in a rigorous way from a realistic
microscopic model, i.~e.~without semiclassical approximations and
phenomenological assumptions. By means of a SU(2) rotation of the fermionic
spinor one can further derive two alternative descriptions of the doped AF
\cite{prb}. In the laboratory reference-frame with a globally fixed SQA, the
doped AF is described by the O(3) non-linear $\sigma$-model for the
order-parameter field $\vec n$ of the spin-background and current-current
interactions between the doped fermions and the order-parameter field.
On the other hand, in the rotating reference-frame with local
SQA defined by the local direction of $\vec n$, the long-wavelength
low-energy theory for the doped AF is given by an U(1) gauge theory
with bosonic and fermionic degrees of freedom minimally coupled to
an U(1) vector gauge field. The present work presents an analysis of the
gauge theory in the rotating reference-frame.
\par
Within a large-$N$ expansion of the U(1) gauge theory we will explicitly
show 1) that there occurs a quantum phase-transition from the
antiferromagnetically ordered N\'eel-state to a QD spin-liquid phase at $T=0$
induced by doping and 2) that the U(1) gauge field mediates a long-range
interaction between the fermionic and bosonic fields, where the strength of
the coupling is given by the spin-gap. Therefore, bound states between
excitations with opposite charge (as defined by the coupling to the
gauge field) are expected to form in the QD phase. The possible
bound states correspond to spin-waves with a gap in the excitation spectrum,
spinless fermions and pairs of fermions. In this frame the phenomena of
charge-spin separation and pairing are intimately connected with the
opening of a spin-gap and have the same physical origin, i.~e.~the
long-range interaction induced by the vector gauge field.
\par
In order to clarify the connection of the gauge theory studied here
with a microscopic model, we summarize in the following the main steps involved
to reach the effective field-theory \cite{prb}.
Our starting point is the strong coupling limit of the three-band
Hubbard model for the CuO$_2$ planes, which is given by the spin-fermion model
\cite{prev} and acts in the subspace of singly occupied Cu-sites:
\begin{eqnarray}
\label{sfm}
\hat H =  \sum_{<j,k>\atop\sigma}
           t_{jk} \: c_{j,\sigma}^\dagger\, c_{k,\sigma}
       + J_K \sum_i \vec R_i \cdot \vec S_i
        +  J_H \sum_{<i,i'>} \vec S_i \cdot \vec S_{i'} \; ,
\end{eqnarray}
where $c_{j,\sigma}^\dagger$ and $c_{j,\sigma}$ create and annihilate holes
with spin-projection $\sigma = \uparrow , \downarrow $ in an oxygen-orbital.
The kinetic term describes hopping processes between oxygen-sites and can
include a direct oxygen-oxygen hopping as well as an effective hopping via
the central copper-site.
Furthermore the Hamiltonian contains a non-local, Kondo-like
interaction between the localized Cu-spins $\vec S_i$ and the
neighboring O-holes
$\vec R_i = \sum_{(j,k;i)\atop\alpha ,\beta}
                 (-1)^{\alpha_{ij}+\alpha_{ik}} \:
c_{j,\alpha}^\dagger\;\vec\sigma_{\alpha\beta}\;c_{k,\beta}$
(here the summation goes only over the indices $j$ and $k$ denoting the
oxygen-sites around the central copper-site $i$), where $\vec \sigma$ are
the Pauli matrices, as well as an antiferromagnetic Heisenberg
superexchange interaction between nearest neighbor Cu-spins.
For a connection of the parameters
of the spin-fermion model with the ones of the
three-band Hubbard model
as well as the definitions of the phase factors $\alpha_{ij}$
which accounts for the d- and p-wave symmetry of the underlying
orbitals we refer to Ref.~\cite{prb}.
\par
In order to construct an effective theory for the long-wavelength low-energy
sector of the spin-fermion model we have to integrate out all the high-energy
modes of the model. For the spin part this is done by means of an expansion
around the adiabatic limit within the method of generalized Berry phases
\cite{prb,antimo}.
The continuum-limit in the order-parameter field $\vec \phi_i$ of the
spin-background is achieved by means of a gradient-expansion around the
AF saddle-point $\vec \phi_i = (-1)^i \vec n_i$, where the lattice constant
$a$ (or the inverse of an equivalent ultraviolet cutoff) serves as
the small expansion parameter. Consequently, the Kondo-like interaction in
the Hamiltonian Eq.~(\ref{sfm}) goes over to essentially the same form but
with the order-parameter $\vec n_i$ instead of the spin-operator $\vec S_i$.
Therefore it becomes a staggered potential for the fermions in the
antiferromagnetic ground-state and together with the hopping parts it
determines the dispersion of the fermions which has minima located at
$\vec k_{min}=(\pm\pi/2 , \pm \pi/2)$.
Since we are interested in the low doping regime and low-energy sector of the
model it is justified to substitute the whole band structure by parabolic bands
located at $\vec k_{min}$.
In this way we obtain the following continuum theory:
\begin{eqnarray}
\label{nlsm}
S_{Nl\sigma}&=&\frac{1}{2g}\int\limits_0^\beta d\tau\:\int d^2r\:
             \left[\Bigl(\partial_r \vec n(\vec r,\tau)\Bigr)^2
             +\frac{1}{c^2}\:\Bigl(\partial_\tau \vec n(\vec r,\tau)\Bigr)^2
             \right] \; ,
\\
S_F&=&
\int_0^\beta d\tau \int d^2r \: \biggl[c^{\dagger} \partial_\tau c
+ \frac{1}{2m_r}\;\partial_r c^{\dagger}
\; \partial_r  c
- \gamma_r\; (\partial_r \vec n)^2 \; c^{\dagger} c
+ \sum_\mu \gamma_\mu \; \vec J_\mu^S \cdot \vec J_\mu^F \biggr]\;\; ; \;\;
\mu = (r,\tau)
\end{eqnarray}
where the first two terms in $S_F$ describe the kinetics of the fermions with
perpendicular $m_\perp$ and parallel $m_\parallel$ masses with respect to the
borders of the magnetic Brillouin-zone. The remaining parts denote the
interaction of the fermions with the antiferromagnetic spin-background, whose
dynamics is controlled by the O(3) non-linear $\sigma$-model with
$g=2\sqrt 2\; a/s$ being the coupling constant and $c=2\sqrt 2\; a J_H s$
being the spin-wave velocity. $s$ denotes here the length of the spin.
The third term of $S_F$ gives a renormalization of the spin-stiffness of the
non-linear $\sigma$-model, whenever
the corresponding site is occupied by a hole. The fourth term
gives current-current couplings similar to the ones obtained
by Shraiman and Siggia for the $t$-$J$ model \cite{shra1}.
For the temporal component we have a
coupling of the local spin-density $\vec J_\tau^F=c^\dagger\vec \sigma c$
of the fermions to the background magnetization
$\vec J_\tau^S = i\vec n\times \partial_\tau \vec n$
whereas the spatial components denotes a coupling between
the spin-current of the holes
$\vec J^F_r= i (\partial_r c^\dagger\vec\sigma c -
c^\dagger\vec\sigma\partial_r c)$
and the magnetization current of the background
$\vec J_r^S= \vec n\times \partial_r \vec n$.
The connections of the corresponding coupling constants as
well as the masses to the microscopic parameters of the spin-fermion model are
given in Ref.~\cite{prb}.
For simplicity we will use isotropic parameters
$\gamma=\gamma_\perp=\gamma_\parallel$ and $m=m_\perp = m_\parallel$
in the following. However, such a situation is close to the case given by a
parameter set for the underlying three-band Hubbard model, which, as quantum
Monte Carlo simulations showed \cite{dopf}, is appropriate for a quantitative
description of various physical quantities of the HTSC.
\par
Alternatively one can describe the system using a reference-frame,
which is defined by a local SQA for the fermionic
spinor in the direction of the order-parameter field $\vec n$.
This change of reference-frame is defined by the following local $SU(2)$
rotation of the fermionic spinor
\begin{eqnarray}
\label{spinor}
p (\vec r,\tau)=U^\dagger (\vec r,\tau)\: c (\vec r,\tau)
\end{eqnarray}
where the $SU(2)$ rotation $U$ must fulfill
$U^\dagger(\vec r,\tau)\;\vec\sigma\cdot\vec n(\vec r,\tau)\;
U(\vec r,\tau)=\sigma^z$.
Using a $CP^1$ representation \cite{cp1} for the rotation matrix $U$
\begin{eqnarray}
\label{cp1}
U= \left( \begin{array}{cc}
               z_1  &  -\bar z_2      \\
               z_2  &   \bar z_1
             \end{array}
      \right)
\end{eqnarray}
with $\bar Z Z = 1$ and $\bar Z = (\bar z_1,\bar z_2)$ denoting a
bosonic spinor, the transition to the rotating reference-frame is
achieved by the following set of equations:
\begin{eqnarray}
\label{trafo}
\mbox{uniform SQA} &\longleftrightarrow& \mbox {rotating SQA} \nonumber \\
c_\mu &\longleftrightarrow& p_\mu \nonumber \\
\partial_\mu - \frac{i}{2}\vec \sigma \cdot
                 \left( \vec n \times \partial_\mu \vec n \right)
 &\longleftrightarrow& \partial_\mu + i A_\mu \sigma^z \\
\frac{i}{2}\vec \sigma \cdot \left( \vec n \times
\partial_\mu \vec n \right)
&\longleftrightarrow&
K_\mu=i B_\mu \sigma^{-} + i \bar B_\mu \sigma^{+}
\nonumber
\end{eqnarray}
Here we have introduced a composite gauge field $A_\mu$ and
denoted the off-diagonal contributions by $B_\mu$ and $\bar B_\mu$:
\begin{eqnarray}
\label{gauge-fields}
A_\mu&=&-i\bar Z\partial_\mu Z  \\
B_\mu      &=& i \left( z_2 \partial_\mu z_1 -
                        z_1 \partial_\mu z_2
                 \right)
            = Z^T \sigma^y \partial_\mu Z     \; , \\
\bar B_\mu &=& i  \left( \bar z_1 \partial_\mu \bar z_2 -
                         \bar z_2 \partial_\mu \bar z_1
                  \right)
            = -\bar Z \sigma^y \partial_\mu \bar Z^T \; ,
\end{eqnarray}
which are related to the $SU(2)$ rotation via
$U^\dagger \partial_\mu U
= i \sigma^z A_\mu + i\bar B_\mu \sigma^+ + i B_\mu \sigma^-$.
As a result of this transformation we get
the following fermionic action
\begin{eqnarray}
\label{local}
S_F=\int_0^\beta d\tau\int d^2r \:
\biggl\{
       p^\dagger \left( D_\tau^F + 2\tilde \gamma_\tau K_\tau \right) p
    +  \frac{1}{2m} \bar D_r^F  p^\dagger \; D_r^F p
%\nonumber \\ &+&
+
\tilde \gamma
       \left[2\left(\partial_r p^\dagger K_r p
                   - p^\dagger K_r \partial_r p
             \right)
             - 4 p^\dagger K_r K_r p
       \right]
\biggr\} \; ,
\end{eqnarray}
with $D_\mu^F = \partial_\mu+i\:\sigma^z A_\mu$, ($\mu \in (\tau,\vec r)$)
denoting a covariant derivative for the fermions and $\tilde
\gamma_\tau = \gamma_\tau+1$ and $\tilde \gamma= \gamma + 1/2m$.
Accordingly the O(3) non-linear $\sigma$-model transforms within this
scheme by applying the Hopf projection \cite{hopf}
defined by
\begin{eqnarray}
\label{hopf}
\vec n = \bar Z \vec \sigma Z \; \; , \; \;
\vec n^2 =1 \; \Leftrightarrow \; \bar Z Z = 1 \; .
\end{eqnarray}
to Eq.~(\ref{nlsm}) into the
$CP^1$ model
\begin{eqnarray}
\label{cp1m}
S_{CP^1}&=&\frac{2}{g}\int_0^\beta d\tau\int d^2r \:
\left\{
       \bar D_r^B \bar Z \; D_r^B Z
       +\frac{1}{c^2}\bar D_\tau^B \bar Z \; D_\tau^B Z
\right\}
\end{eqnarray}
where $D_\mu^B = \partial_\mu -i A_\mu$ denotes a covariant derivative
for the bosons. Here and in the rest of the paper the superscripts $F$ and $B$
refer to the contributions of the bosons and the fermions, respectively.
The whole action in the rotating reference-frame $S=S_F+S_{CP^1}$ is invariant
under a $U(1)$ gauge transformation.
%At this stage we want to point out that
%here the gauge fields arise by relating an $SU(2)$ rotation in spin-space and
%a vector on the sphere $S^2$. The manifold $SU(2)$ is isomorphic to $S^3$,
%however, the vector $\vec{n}$ in $S^2$ fixes only two of the three angles in
%$S^3$, and hence, a phase remains free. In contrast to this, the gauge fields
%in the slave-boson approach of the $t-J$ model \cite{naga} arise from
%implementing the constraint of forbidden double-occupancy in the action by
%introducing a lagrange multiplier whose phase-fluctuations are identified as
%the gauge field in the long-wavelength limit. Furthermore, the fermionic
%fields in Eq.~(\ref{local}) correspond to the bare holes introduced by doping.
\par
{}From the definitions of covariant derivatives we get the following
charges of the bosons and fermions with respect to the gauge
field $A_\mu$:
\begin{equation}
\label{charges}
\begin{array}{|r|c|}\hline
             &  \mbox{charge}  \\ \hline
      z_1    &  +1                \\
      z_2    &  +1                 \\
p_\uparrow   &  +1                 \\
p_\downarrow &  -1                \\ \hline
\end{array}
\end{equation}
The gauge theory defined by Eqs.~(\ref{local}) and (\ref{cp1m}) is the
starting point for the considerations in the present work.
\par
The outline of the paper is as follows. Section \ref{largeN} is devoted to a
large-$N$ expansion of the U(1) gauge theory, where we will show 1) that the
theory describes a doping induced quantum phase transition from an
antiferromagnetically ordered state to a quantum disordered spin-liquid state
(Sec.\ref{RC to QD})
and 2) that the transverse gauge fields that mediate a long-range interaction
between the bosons and fermions has a strength given by the mass of the
$Z$-bosons, i.e.\ the spin-gap (Sec.\ref{propagator of the gauge field}).
In Section \ref{discussion} we will discuss the
physical consequences of the gauge field and consider the connection between
the opening of a spin-gap and charge-spin separation and pairing.
In this section we will also address the implications of the bound states
on various physical properties of the cuprates in the low-doping regime and
we will give a brief summary of our main results and conclusions.
A short account of the results above is given in Ref.~\cite{epl}.

\section{Large-$N$ expansion for the gauge theory}
\label{largeN}

The gauge field $A_\mu$ is not an independent dynamical degree of freedom
since it is composed by two $Z$-bosons and has no kinetic part in the
action (\ref{local}). Nevertheless, as usual \cite{dadda}, such a kinetic term
will arise due to the fluctuations of the bosonic as well as of the
fermionic components. The calculation of such fluctuations is, however, not
straightforward since the coupling constant for the gauge fields is of order
unity, implying that there exists no small expansion parameter in the theory.
In order to overcome these difficulties a large-$N$ expansion is implemented,
where $N$ is the number of bosonic or fermionic flavors \cite{dadda}.
\par
Within the large-$N$ expansion it will be shown that doping drives the system
from an antiferromagnetically ordered state at $T=0$ to a quantum disordered
spin-liquid state.  A mass for the $Z$-bosons is generated dynamically, such
that a "spin-gap" enters naturally in the description of the low-energy
properties of the system, much in the same way as in recent analysis of the
$O(3)$ non-linear $\sigma$-model \cite{chn,sach}.
In the present case, however, the
non-diagonal contributions $\propto K_\mu$ will lead to a doping dependence
of the mass and, hence, a doping induced quantum phase-transition takes place.
In the course of the large-$N$ expansion also the dynamics of the gauge fields
is obtained, where the new energy-scale generated dynamically (i.e.\ the mass
for the $Z$-bosons) determines the strength of the propagator of the
gauge fields. This fact reveals an intimate connection between the spin-gap
and the properties of the charge-carriers in the QD phase.
\par
As a first step we introduce $L$ copies of fermions with spin $\uparrow$ and
with spin $\downarrow$
\begin{eqnarray}
(p_\uparrow,p_\downarrow)\longrightarrow
(p_\uparrow^1,...,p_\uparrow^L ; p_\downarrow^1,...,p_\downarrow^L)\; .
\end{eqnarray}
Accordingly the $CP^1$ fields are generalized to $CP^{N-1}$ ones
\begin{eqnarray}
\bar Z = (\bar z_1,\bar z_2)
\longrightarrow \bar Z = (\bar z_1,...,\bar z_N)
\end{eqnarray}
with $N=2L$ so that the ratio of fermionic and bosonic degrees of freedom
remains the same after this generalization.
\par
We proceed further by
introducing  sources for the fields and obtain
(after proper renormalization of the fields) the following generating
functional
\begin{eqnarray}
\label{functional1}
& &{\cal Z} [J , \bar J , \eta , \bar \eta , Q_\mu] =
\int{\cal D} Z {\cal D} \bar Z {\cal D} p {\cal D} p^\dagger  \:
\prod_{\vec r,\tau} \: \delta \left(\bar Z Z - \frac{2N}{g}\right)
\nonumber \\
& &\cdot \exp \left\{
-S_F - S_{CP^{N-1}} +\int_0^\beta d\tau \int d^2r
     \left(
       \bar \eta p+ p^\dagger \eta + \bar Z J + \bar J Z +
       \frac{g}{2N} Q_\mu A_\mu
     \right)
\right\} \; .
\end{eqnarray}
Since the action $S_F$ is bilinear in the fermionic variables they can
easily be integrated out:
\begin{eqnarray}
S_F& = & - N\mbox{Tr}\ln\Delta_F
     =   - N\mbox{Tr}\ln(\Delta_F^0+\Delta_F^A+\Delta_F^K)\; .
\end{eqnarray}
Here we have divided the fermionic contribution into three terms
\begin{eqnarray}
\Delta_F^0       &=& -\frac{1}{2m} \partial^2_r +2g \partial_\tau \; ,\\
\Delta_F^A &=& -\frac{1}{2m}
%\left(\frac{g}{2N}\right)
      \left[  i \sigma^z 2 \frac{\delta}{\delta Q_r}\partial_r
            + i \sigma^z\partial_r \left(\frac{\delta}{\delta Q_r}\right)
      \right]
    + 2g \; i\sigma^z \frac{\delta}{\delta Q_\tau}
%\left(\frac{g}{2N}\right)
    + \frac{1}{2m} \frac{\delta^2}{\delta Q_r^2}
%\left(\frac{g}{2N}\right)^2
\; ,\\
\Delta_F^K &=& i \tilde \gamma_\tau \left(\frac{g}{2N}\right)
               2 \left(B_\tau\sigma^- + \bar B_\tau\sigma^+\right)
               + \tilde \gamma  \left(\frac{g}{2N}\right)^2 4 B_r \bar B_r
                 \nonumber \\
           &+&     \tilde \gamma 2 \left(\frac{g}{2N}\right)
                 \left[2\left(B_r \sigma^- + \bar B_r \sigma^+\right)\partial_r
                   + \partial_r\left(B_r \sigma^- + \bar B_r \sigma^+\right)
                   \right] \; ,
\end{eqnarray}
the first one is the free kinetic part, the second one contains the
contributions from the gauge field $A_\mu$, which we
have substituted by a functional derivative with respect
to its source
%($\frac{g}{2N} A_\mu \rightarrow \frac{\delta}{\delta Q_\mu}$)
($(g/2N)\; A_\mu \rightarrow \delta/\delta Q_\mu$)
%$Q_\mu$
as is clear from Eq.~(\ref{functional1}),
and the third term contains the off-diagonal
contributions $K_\mu$.
This division turns out to be useful, since
due to the underlying $SU(2)$ structure
the $A_\mu$ and $K_\mu$-terms in $S_F$ do not mix within the
large-$N$ expansion and we can therefore treat them
independently form each other.
\par
We first show that the $K_\mu$-terms modify the $CP^{N-1}$ model.
An expansion of the non-diagonal part in powers of $1/\sqrt{N}$ yields:
\begin{eqnarray}
\label{bterme1}
S_F^{K}&=& \int d^{3}\!p\: N \left(\frac{g}{2N}\right)^2
4 \bar B_r(p) B_s(-p)
\biggl[ \tilde \gamma \delta_{rs} \int d^{3}\!q\: G_F(q)
+\frac{1}{2} \tilde \gamma^2
\int d^{3}\!q\:  G_F(q) (2q+p)_r (2q+p)_s G_F(q+p)
\biggr] \nonumber \\
&+&\int d^{3}\!p\:  N \left(\frac{g}{2N}\right)^2
4 \bar B_\tau (p) B_\tau (-p)
\frac{\tilde \gamma_\tau^2}{2} \int d^{3}\!q\: G_F(q) G_F(q+p) \; ,
\end{eqnarray}
where $G_F(q)= 1/(i\nu_n-\vec q^2/2m)$
is the fermionic Greens function
and where we have introduced the short hand notation
$\qint = 1/\beta \sum_{\nu_m} \int d^2q/(2\pi)^2$.
Evaluating the integrals in Eq.~(\ref{bterme1}) in the limit
$\vec p\rightarrow 0$ and $\omega/\vert\vec p\vert \rightarrow 0$
(after proper analytic continuation) and noticing that
$(g/2N)\bar B_\mu B_\nu = {\bar D}_\mu^B \bar Z\: D_\nu^B Z$
one gets for $S_F^{K}$ together with the pure bosonic
contribution the following generalized $CP^{N-1}$ model:
\begin{eqnarray}
\label{bterme2}
S^{\mbox{gen.}}_{CP^{N-1}}=S_{CP^{N-1}}+ S_{F}^{K}
=\int_0^\beta d\tau \int d^2r \:
\left\{ f_r^\delta{\bar D}^B_r \bar Z\; D_r^B Z
+\frac{1}{c^2} f_\tau^\delta
{\bar D}^B_\tau \bar Z\; D_\tau^B Z
\right\}
\end{eqnarray}
with doping dependent coupling constants
\begin{eqnarray}
f_r^\delta   &=& 1+2g\:\delta\left(\tilde \gamma-2m\:\tilde {\gamma}^2\right)
              =  1-2g\delta\left(1+2\gamma\:m\right)\gamma \; , \\
f_\tau^\delta &=&  1+2g c^2 \tilde \gamma_\tau^2 \frac{m}{4\pi}\Theta (\delta)
\; ,
\end{eqnarray}
where $\Theta (\delta)=1 $ for $\delta>0$ and zero otherwise.
Before we can integrate over the bosonic variables in the path integral,
we have to decouple the quartic terms. This is achieved by means of the
following Hubbard-Stratonovich transformation
\begin{eqnarray}
& &\exp \left\{
\frac{g/f_r^\delta}{2N}\int_0^\beta d\tau\int d^2r
\left[ -(\bar Z \partial_\mu Z)
(\bar Z \partial_\mu Z)+ i Q_\mu (\bar Z \partial_\mu Z)\right]
\right\} \nonumber \\
& & =\int {\cal D} \lambda_\mu \exp \left\{
\int_0^\beta d\tau\int d^2r
\left[ -\frac{2}{g/f_r^\delta} \lambda_\mu \lambda_\mu +
\frac{2i}{\sqrt{N}} \lambda_\mu (\bar Z \partial_\mu Z) +
\frac{1}{\sqrt{N}} \lambda_\mu Q_\mu -\frac{g/f_r^\delta}{8N}
Q_\mu Q_\mu \right]
\right\} \; ,
\end{eqnarray}
where we have rescaled the bosonic fields in such a way that now
the constraint reads $\delta \left(\bar Z Z - \frac{2N}{g/f_r^\delta}\right)$
and we have absorbed the factor $f_\tau^\delta/c^2$ in a redefinition of the
inverse temperature $(c/\sqrt{f^\delta_\tau}) \beta\rightarrow \beta$.
Since the introduced Hubbard-Stratonovich
field $\lambda_\mu$ couples now linearly to the source $Q_\mu$
it plays from now on the role of the gauge field.
Furthermore we include the constraint for the bosonic variables
via a lagrange multiplier field $\alpha$ into the action:
\begin{eqnarray}
\delta\left(\vert z\vert^2-\frac{2N}{g/f_r^\delta}\right)=
\int {\cal{D}}\alpha\exp\biggl\{\int_0^\beta d\tau\int d^2r\left[
\frac{i}{\sqrt{N}}\alpha\left(\bar{z}z-\frac{2N}{g/f_r^\delta}\right)\right]
\biggr\}\; .
\end{eqnarray}
After these modifications the action is bilinear in the bosonic variables
and they can be integrated out.
In this way we get the following effective action
\begin{eqnarray}
\label{effective}
S_{\mbox{eff}}=N\mbox{Tr} \ln \Delta_B - N \mbox{Tr} \ln
\left(\Delta_F^0 + \Delta_F^A \right)
+ i\frac{2\sqrt{N}}{g/f_r^\delta}
  \int_0^\beta d\tau\int d^2r \:\alpha (\vec r,\tau)\; ,
\end{eqnarray}
with
\begin{eqnarray}
\Delta_B &=& -D^B_\mu D^B_\mu +M^2 -\frac{i\alpha}{{\sqrt{N}}}\; ,
\\
D_\mu^B  &=& \partial_\mu + \frac{i}{{\sqrt{N}}} \lambda_\mu \; .
\end{eqnarray}
Please notice that after the steps following Eq.~(\ref{bterme2})
the contributions of the non-diagonal $K_\mu$-terms are now
contained in the doping dependent coupling constant in front of the last term
of Eq.~(\ref{effective}).
At this stage we have also introduced a mass $M$ for the bosonic variables
which is arbitrary since the corresponding mass term $M^2 \bar Z Z$
gives only a constant in the
path integral due to the constraint for the bosons \cite{dadda}.
Resubstituting the functional derivative with
respect to the source field $Q_\mu$ in the fermionic part
by the gauge field $\lambda_\mu$ and expanding
the remaining fermionic part and the bosonic one
in powers of $1/\sqrt{N}$ leads to
\begin{eqnarray}
\label{1/N}
S_{\mbox{eff}} = \sum_{\nu = 1}^\infty N^{1-\frac{\nu}{2}} S^{(\nu)}\; .
\end{eqnarray}
In the limit of large $N$ only the first two terms will survive. The
first term is given by
\begin{eqnarray}
\label{s1}
S_{(1)}=i{\alpha}(\vec p=0,\omega=0)
\left( \frac{2}{g/f_r^\delta}-
       \qint \frac{1}{\vec q^2+\nu^2+M^2}
\right) \; ,
\end{eqnarray}
where $\nu = 2 \pi m/\beta$ are bosonic Matsubara frequencies.
We want to stress here that without taking into account the
non-diagonal terms $\propto K_\mu$ there would appear no doping dependent
parameter as $f_r^\delta$ in Eq.~(\ref{s1}) since the diagonal fermionic
contributions to $S^{(1)}$ are identical to zero due to the vanishing trace of
the Pauli matrix $\sigma^z$.
The second important term in the large $N$-expansion is given by the
polarization insertions of the fermions and bosons for the gauge field
$\lambda_\mu$ as well as for the constraint field $\alpha$:
\begin{eqnarray}
\label{s2}
S^{(2)} = \frac{1}{2} \pint
\left\{ \lambda_\mu (\vec p,\omega_n)\;
 \Bigl(\Pi_{\mu\nu}^B (\vec p,\omega_n) + \Pi_{\mu\nu}^F(\vec p,\omega_n)\Bigr)
        \lambda_\nu (-\vec p,-\omega_n)
- \alpha (\vec p,\omega_n)\; F(\vec p,\omega_n)\;
              \alpha (-\vec p,-\omega_n)
\right\} \; ,
\end{eqnarray}
where
the polarization tensors
are defined as follows:
\begin{eqnarray}
\label{bosons}
\Pi^B_{\mu\nu}(\vec p,\omega_n)&=&2\delta_{\mu\nu}
\qint \frac{1}{\nu_m^2+{\vec q}^2+M^2}
%\\ \nonumber
- \qint \frac{(2q_\mu+p_\mu)(2q_\nu+p_\nu)}
{\Bigl[\vec q^2+\nu_m^2+M^2\Bigr]\Bigl[(\vec q+\vec p)^2
+(\nu_m+\omega_n)^2+M^2\Bigr]}\; ,
\\
\label{fermions}
\Pi_{\mu\nu}^F(\vec p,\omega_n)&=&\delta_{\mu r}
\biggl\{
2\delta_{r s}\frac{1}{2m}
\qint
\frac{1}{i\nu_m - \frac{1}{2m}\vec p^2}
-\left(\frac{1}{2m}\right)^2
\qint
\frac{(2p+q)_r\; (2p+q)_s}{[i\nu_m - \frac{1}{2m}\vec p^2]
[i(\nu_m+\omega_n) - \frac{1}{2m}(\vec p+\vec q)^2]}
\biggr\}
\delta_{s \nu}
\nonumber \\
&-&
\delta_{\mu 0}
\biggl\{
\frac{1}{2} \qint
\frac{1}{[i\nu_m - \frac{1}{2m}\vec p^2][i(\nu_m+\omega_n) -
\frac{1}{2m}(\vec p+\vec q)^2]}
\biggr\}
\delta_{0 \nu}
\nonumber \\
&-&i \frac{1}{2} \left(\frac{1}{2m}\right)
\delta_{\mu 0}
\biggl\{
\qint
\frac{(2p+q)_r}{[i\nu_m - \frac{1}{2m}\vec p^2][i(\nu_m+\omega_n) -
\frac{1}{2m}(\vec p+\vec q)^2]}
\biggr\}
\delta_{r \nu}  \; ,
\\
\label{flagrange}
F(\vec p,\omega_n)&=&
\qint \frac{1}{\vec q^2+\nu_m^2+M^2}\;
\frac{1}{(\vec q+\vec p)^2+(\nu_m +\omega_n)^2+M^2} \; .
\end{eqnarray}
As a consequence of the local $U(1)$ gauge-symmetry the polarization tensor
$\Pi_{\mu\nu}= \Pi_{\mu\nu}^F + \Pi_{\mu\nu}^B$ has to
be transversal $p^\mu \Pi_{\mu\nu} = 0 $ and this property turns out to be
very useful in evaluating the polarization tensors by means of a
tensorial decomposition.
%The important point here is that due to fluctuations of the fermions and
%%bosons
%a kinetic energy term for the gauge field $\lambda$ was generated dynamically,
%indicating that from now on the gauge field $\lambda_\mu$ can be considered as
%an own dynamically degree of freedom.

\subsection{From the renormalized classical to the quantum disordered
phase by doping}
\label{RC to QD}

\label{phase diagram}

Since in the limit $N \rightarrow \infty$ the expression
$\exp\{-\sqrt{N}S^{(1)}\}$ gives a strongly fluctuating
factor in the integrand of the generating functional (\ref{functional1})
only the contribution $S^{(1)}\equiv 0 $ will survive. Evaluating the integral
in the expression (\ref{s1})
this saddle point equation establishes a definite relationship between
the mass $M$ of the $Z$-bosons, the coupling constant $g$, the
temperature $T$, and the doping $\delta$:
\begin{eqnarray}
\label{mass}
M=\frac{2}{\beta}\mbox{arcsinh}
\left[ \frac{1}{2}\exp
\left\{-4\pi\beta
\left(\frac{f_r^\delta}{g}-\frac{1}{g_c}\right)
\right\}
\right]
\end{eqnarray}
with $g_c=8\pi/\Lambda$ denoting the critical coupling constant which
depends on the ultraviolet cut-off $\Lambda$.
Eq.~(\ref{mass}) shows that the previously arbitrary mass $M$
is now fixed by the saddle point condition and is therefore
dynamically generated \cite{dadda}.
By introducing the renormalized spin stiffness
$\rho=2/g - 2/g_c$ at zero doping, where the system is
in the N\'eel ordered phase (at $T=0$) \cite{chn},
we get the following expression for the mass $M$
which is essentially the inverse of the spin-spin correlation length $\xi$
\begin{eqnarray}
\label{mass with doping}
M=\xi^{-1}=
\frac{2}{\beta}\mbox{arcsinh}\left[\frac{1}{2}
\exp
\left\{-2\pi\beta
     \Bigl(\rho - 4\gamma(1+2\gamma \; m)\delta \Bigr)
\right\}\right]\; .
\end{eqnarray}
According to Chakravarty, Halperin and Nelson (CHN) \cite{chn}
we have to distinguish three different regimes depending on the value
of the argument
$y=\frac{2\pi}{T} \left(\rho - 4\gamma(1+2\gamma \; m)\delta \right)$
in the exponential of Eq.~(\ref{mass with doping}).
For $y\gg 1$ the system is in the
renormalized classical (RC) regime where the spin-spin correlation length $\xi$
diverges exponentially as $T$ tends to zero, indicating a ground state
with long-range antiferromagnetic N\'eel order at $T=0$ ($\beta = \infty$)
\begin{eqnarray}\label{geordnet}
\xi =\beta \exp\left[2\pi\beta
\Bigl(\rho - 4\gamma(1+2\gamma \; m)\delta\Bigr)\right] \; .
\end{eqnarray}
Contrary for $y \ll -1$ the correlation length
tends to a constant value at $T=0$
\begin{eqnarray}\label{ungeordnet}
\xi =\frac{1}{4\pi[4\gamma(1+2\gamma \; m)\delta-\rho]} \; ,
\end{eqnarray}
implying that the system is in a QD spin-liquid state.
{}From the above discussion it is clear that there must be a phase transition
due to doping at $T=0$ where the system goes over from the N\'eel state to the
QD spin liquid state.  The corresponding critical doping $\delta_c$, where
this quantum phase transition occurs, is defined by:
\begin{eqnarray}
\label{critical doping}
\delta_c =\frac{\rho}{4\gamma(1+2\gamma\;m)} \; .
\end{eqnarray}
Thus the critical doping $\delta_c$ is completely determined by the
experimental value of the spin-stiffness $\rho$ at zero doping and the
microscopic parameters $m$ and $\gamma$.
For an estimate of the critical doping $\delta_c$ we use
a parameter set \cite{parameter} of the underlying
three-band Hubbard model, which lies in the strong coupling regime,
where the spin-fermion model applies.
Also this parameter set is very realistic as was shown by
quantum Monte Carlo simulations \cite{dopf} through a comparison of
numerical results and experimental data for the cuprates.
Furthermore the assumption of isotropic masses and coupling constants
in the theory made in the introduction is justified using this parameter set.
Using the expressions for the mass $m$ and the coupling constant $\gamma$
given in Ref.~\cite{prb} we get $m=3/2$ and $\gamma=1/12$ \cite{parameter}.
With the experimental value $\rho = 1.7\cdot 10^{-2}$ \cite{rho}
we explicitly get $\delta_c\sim 4\%$ \cite{epl}.\par
Finally for $\vert y\vert \ll 1$ the system is in the quantum
critical (QC) regime which is completely determined by the
critical point $(\delta_c,T=0)$ \cite{sach}.

\subsection{Propagator of the gauge field}
\label{propagator of the gauge field}

After we have discussed the phase diagram of the doped AF
we examine here the second term of the large-$N$ expansion,
which consists of the polarization insertions of the
fermions and bosons for
the gauge field $\lambda_\mu$ and for the constraint field $\alpha$.
As already stated, the polarization tensor (\ref{s2}) of the
gauge field $\lambda_\mu$ has to be transverse and therefore the
following tensorial decomposition is useful \cite{mavro}:
\begin{eqnarray}
\label{decomposition}
\Pi_{\mu\nu}(p) = A_{\mu\nu}(p) \: \left[ \Gamma_1^F(p) + \Gamma_1^B(p) \right]
                + B_{\mu\nu}(p) \: \left[ \Gamma_2^F(p) + \Gamma_2^B(p) \right]
                  \; ,
\end{eqnarray}
with $p=(\vec p , \omega_n)$ denoting the wave-vector $\vec p$
together with the Matsubara frequency $\omega_n$.
Here we have defined the two transverse tensors
($p^2 = \vec p^2 + \omega_n^2$)
\begin{eqnarray}
A_{\mu\nu} = \left( \delta_{\mu 0} - \frac{p_\mu p_0}{p^2} \right)
             \frac{p^2}{\vec p^2}
             \left( \delta_{0 \nu} - \frac{p_0 p_\nu}{p^2} \right)
\;\; ; \;\;
B_{\mu\nu} = \delta_{\mu i}
           \left( \delta_{ij} - \frac{p_i p_j}{\vec p^2} \right)
             \delta_{j \nu}
\end{eqnarray}
with the following properties:
\begin{eqnarray}
\label{properties}
A_{\mu\nu}+B_{\mu\nu}=
\delta_{\mu\nu}-\frac{p_\mu p_\nu}{p^2}\; ; \; A_{\mu\nu}B^{\mu\nu}=0\; .
\end{eqnarray}
The scalar functions $\Gamma_1^{F,B} (p)$ and $\Gamma_2^{F,B} (p)$
are evaluated
in Appendix \ref{tensorial decomposition} and are given by:
\begin{eqnarray}
\label{gammab}
\Gamma^B(p) &:=&\Gamma_1^B (p)=
\Gamma_2^B (p) = \frac{p^2+4M^2}{2}F(p) - \frac{M}{4\pi} \; ,
\\
\Gamma_1^F(p) &=& \frac{p^2}{\vec p^2}
%                  \frac{1}{\beta}\sum_{\nu_m}\int \frac{d^2q}{(2\pi)^2}
                  \qint
                  \frac{1}{\left[\frac{1}{2m}\vec q^{\:2} - i\nu_m\right]
                           \left[\frac{1}{2m}\left(\vec q + \vec p\right)^2
                                 -i (\nu_m +\omega_n)\right]}\; ,
\\
\Gamma_2^F(p) &=&  \frac{\rho_F}{m}
                  -\left(\frac{1}{2m}\right)^2 \qint
                    \frac{(2\vec q+\vec p) \cdot (2\vec q+\vec p)}
                         {\left[\frac{1}{2m}\vec q^{\: 2}-i\nu_m\right]
                          \left[\frac{1}{2m}\left(\vec q+\vec p\right)^{\: 2}
                          -i \left(\nu_m+\omega_n\right)\right]}
\nonumber \\
              &+&\left(\frac{1}{2m}\right)^2 \frac{1}{\vec p^2} \qint
      \frac{\left(2\vec q + \vec p\right)\cdot \vec p \hspace*{05mm}
            \left(2\vec q + \vec p\right)\cdot \vec p}
           {\left[\frac{1}{2m}\vec q^{\: 2}-i\nu_m\right]
            \left[\frac{1}{2m}\left(\vec q+\vec p\right)^{\: 2}-i
                              \left(\nu_m+\omega_n\right)\right]}\; .
\end{eqnarray}
After an analytical continuation to real frequencies
$i\omega_n \rightarrow \omega+i0^+$ and evaluating the various integrals
at $T=0$ we get the following expressions for the scalar $\Gamma$ functions
(The details of this calculation are delegated to Appendix \ref{integrals}):
\begin{eqnarray}
\Gamma^B (p) &=& \frac{\vec p^2 - \omega^2 + 4M^2}{8\pi
\sqrt{\vec p^2-\omega^2} }
\arcsin\left(\sqrt{\frac{\vec p^2-\omega^2}{4M^2+(\vec p^2-\omega^2)}}\right)
 - \frac{M}{4\pi} \; ,
\\
\Gamma^F_1 (p)&=& \frac{p^2}{\vec p^2} \frac{m}{2\pi}
          \left( 1 - \frac{x\Theta(x^2-1)} {\sqrt{x^2-1}}
                 +i \frac{x\Theta(1-x^2)} {\sqrt{1-x^2}}
          \right) \; ,
\\
\Gamma^F_2 (p)&=& \frac{\rho_F}{m}
\left[x^2 +   (x-x^3) \frac{\Theta(x^2-1)}{\sqrt{x^2-1}}
          - i (x-x^3) \frac{\Theta(1-x^2)}{\sqrt{1-x^2}}
\right] \; ,
\end{eqnarray}
where $x=|\frac{\omega}{\vec p v_F}|$.
$\rho_F$ is the fermionic density and $v_F$ the Fermi velocity. Within
the Coulomb gauge, $\vec \nabla\cdot\vec \lambda=0$,
we get for the propagator of the gauge field $\lambda_\mu$
\begin{eqnarray}
\label{propagator}
D^\lambda_{00} &=& \frac{p^2}{\vec p^2}
            \left(
              \Gamma^B (\vec p,\omega) + \Gamma^F_1 (\vec p,\omega)
            \right) ^{-1} \; ,    \\
D^\lambda_{0i} &=& 0 \; ,                   \\
D^\lambda_{ij} &=& \left( \delta_{ij} - \frac{p_i p_j}{\vec p^2} \right)
            \left(
              \Gamma^B (\vec p,\omega) + \Gamma^F_2 (\vec p,\omega)
            \right) ^{-1} \; .
\end{eqnarray}
Since we are interested on the long-wavelength low-frequency physics
of the gauge theory it is appropriate to study the limit
$\vec p\rightarrow 0$, $\omega \rightarrow 0$.
Because of the step functions in the fermionic contributions we have
to distinguish two different regions in the $\omega$-$p$ plane. \par
In the region, where $x > 1$, we get
\begin{eqnarray}
\label{lambda}
D^{\lambda^>}_{\tau \tau} &=&\frac{\omega^2}{\vec p^2}
\frac{-M}{\frac{1}{24\pi}\omega^2 + \frac{\rho_F M}{m}} \; ,
\\
\label{spatial}
D^{\lambda^>}_{ij} &=& \left(\delta_{ij}-\frac{p_i p_j}{\vec p^2}\right)
\frac{-M}{\frac{1}{24\pi}\omega^2 - \frac{\rho_F M}{2m}}
\end{eqnarray}
and for $x<1$ we get:
\begin{eqnarray}
D^{\lambda^<}_{\tau \tau}
&=& \frac{1}{\frac{1}{24\pi M}\vec p^2 + \frac{m}{2\pi}} \; ,
\\
\label{propagatorraum}
D^{\lambda^<}_{ij}(\vec p , \omega ) &=&
\left(\delta_{ij}-\frac{p_i p_j}{\vec p^2}\right) \frac{M}{\frac{1}{24 \pi}
{\vec p}^{2} - i \rho_F\frac{M}{m}\left|\frac{\omega}{\vec p v_F}\right|}\; .
\end{eqnarray}
The propagator of the constraint field $\alpha$ is completely
determined by the bosonic polarization and is therefore
independent of the value of $x$
\begin{eqnarray}
\label{alpha}
D^\alpha=F_B^{-1}=8\pi M \; .
\end{eqnarray}
As we see from Eqs.~(\ref{lambda}) - (\ref{alpha}) the time component of the
gauge field propagator as well as the propagator of the Lagrange-multiplier
field $\alpha$  are massive, since they are Debeye screened, and thus
mediate only short-range density-density interactions
between the fermions and bosons.
Here we want to remark that the pole in the
prefactor $\frac{\omega^2}{\vec p^2}$ in Eq.~(\ref{lambda})
is not physical, since it is a consequence of the choosen Coulomb gauge.
In fact one can explicitly see in the axial or in the Lorentz gauge,
that the singularity due to that prefactor in Eq.~(\ref{lambda}) is absent.
Physical poles of the gauge field propagator are only given by the zeros of the
gauge independent combinations
of the fermionic and bosonic scalar functions
$\left( \Gamma^B (\vec p,\omega) + \Gamma^F_1 (\vec p,\omega) \right)$
and $\left( \Gamma^B (\vec p,\omega) + \Gamma^F_2 (\vec p,\omega) \right)$
appearing in the corresponding expressions for the propagator.
\par
In the region $|\frac{\omega}{\vec p v_F}|>1$
the spatial component of the gauge field propagator
has a pole at a finite frequency $\omega^2 = 24 \pi \frac{\rho_F M}{m}$
with a form reminiscent of the plasmon pole in an electron gas.
%with mass $m$ and charge $e$.
Furthermore, in the opposite limit, $|\frac{\omega}{\vec p v_F}|<1$,
the gauge field propagator turns out to be massless
and describes an overdamped mode with a
dispersion law $\omega \propto i p^3$, whose physical origin
can be traced back to the Landau damping
$\propto i \omega/p$ due to fermions near the Fermi surface.
As a consequence of this massless mode the fermions and bosons
experience an effective long-range interaction by exchanging
transverse gauge bosons.
The form of the propagator in Eq.(\ref{propagatorraum}) is
essentially the same as found by Nagaosa and Lee \cite{naga} and Reizer
\cite{reizer}, with a central difference given by the fact that the
mass $M$ of the magnetic excitations
determines the strength of the propagator.
Hence as a special feature of the present theory a new energy scale,
the spin-gap $\Delta \propto M$ measured in neutron scattering
experiments \cite{barz}, enters in a natural way and appears as the
strength of the interaction.
Therefore, we expect that the long-range current-current interaction
mediated by the transverse gauge field becomes only effective in the QD
phase, where the spin-gap $\Delta$
is large compared to the thermal energy $k_BT$.
This implies that the tendency to form bound states only occurs in the
disordered magnetic state, such that the phenomena of charge-spin separation
and pairing in this theory should be restricted to that region.

\section{Discussion and Summary}
\label{discussion}

As was noticed first by Holstein et al.~\cite{holstein} and later by
Reizer \cite{reizer}, the coupling of fermions to a transverse vector gauge
field leads to deviations from the Fermi-liquid picture. However,
in the case of a conventional metal interacting with the electromagnetic
field, the effects of the exchange of transverse gauge bosons are in practice
unobservable since they are suppressed by the ratio of the Fermi-velocity and
the speed of light $\left(\frac{v_F}{c}\right)^2$. In contrast to those
systems, in the above gauge theory for the doped AF the coupling constant is of
order unity and therefore no such small parameter enters, a feature common to
gauge theories developed recently for strongly correlated metals
\cite{naga,polchinski,blok}.
Calculating the self-energy for the fermions at $\vec k_F$
in lowest order of perturbation theory one gets an unusual frequency dependence
$\Sigma \propto \omega^{\frac{2}{3}}$, which implies that the quasi-particle
weight $Z= \left| 1- \frac{\partial}{\partial\omega} \mbox{Re}\:
\Sigma_{\omega\rightarrow 0} \right|^{-1} $ vanishes at the Fermi surface
signaling a break down of Fermi-liquid picture \cite{naga}, and hence,
a perturbative expansion in the coupling constant is in this case not valid
\cite{blok} and one has to implement some non-perturbative approaches.
In spite of the conflicting claims in the recent literature concerning the
one-particle properties of gauge theories for strongly correlated metals
\cite{stamp}, we would like to point out that the results
obtained there are not directly applicable to our case, since the physically
relevant one-particle properties appear only after a rotation back to the
global refence frame. This amounts in the lowest order of the large
$N$-expansion to a convolution of the free fermionic and bosonic propagators
and the corrections due to the gauge fields are obtained at ${\cal O} (1/N)$.
Results pertinent to this point will be presented elsewhere \cite{felix}.
An even stronger difference with the general discussion of gauge theories in
this context should be expected due to the possibility of formation of
bound states, that we discuss in the following.
\par
The singular mode of the gauge field mediates a long-range current-current
interaction between the $Z$-bosons and the fermions, which in the static
limit reduces to a logarithmic potential. Moreover since the strength of
the singular gauge field propagator is proportional to the spin-gap $\Delta$
we expect the formation of bound states only in the QD phase, where the
spin-gap is large compared to the thermal energy $k_B T$.
Hence particles with opposite charge will bind and the
low-energy sector of the physical
spectrum of the theory will contain only states with zero charge with
respect to the gauge field $\lambda$. Since the charge of the fields
is given by their spin-projection, the
bound states correspond to spin-singlets.
According to table (\ref{charges}) the following three
bound states are possible:
\begin{enumerate}
\renewcommand{\labelenumi}{(\alph{enumi})}
\item Bound states between two bosons:
      \begin{eqnarray}
      \label{boson-boson}
      \bar Z  - Z
      \end{eqnarray}
      These bound states correspond to spin-waves excitations (remember
      $\vec n = \bar Z \vec \sigma Z$) at the
      antiferromagnetic wave vector $\vec Q=(\pi , \pi)$
      with a gap $\Delta \propto M$ in the spectrum.
\item Bound states between a fermion and a boson:
      \begin{eqnarray}
      \label{boson-fermion}
       Z  - p
      \end{eqnarray}
      These bound states have an electric charge $+e$ but have no spin.
      Thus they describe spinless charge-excitation with fermionic
      character.
      Thus, this scenario gives an alternative way to charge-spin separation,
      where the bare excitations are just spin-$\frac{1}{2}$ fermions but the
      renormalized ones are spinless.
\item Bound states between two fermions:
      \begin{eqnarray}
      \label{fermion-fermion}
            p_\downarrow  - p_\uparrow
      \end{eqnarray}
      Such a bound state describes the pairing of two fermions in a
      spin-singlet with charge $2e$.
      Hence, charge-spin
      separation and pairing are intimately connected
      in our case and result from the same interaction.
\end{enumerate}
A similar description for holes in a spin-liquid state was derived by Wen
\cite{wen1} starting with a one-band Hubbard model.
However, in that case
charge and spin degrees of freedom are separated from the very beginning
(the bare fermions do not carry spin) and hence
the phenomenon of charge-spin separation
does not follow from the formation of bound states as in our theory.
Quite on the contrary, the bound states between fermions and bosons
in that theory carry charge $e$ and spin, implying that
charge-spin separation only occurs when there is no
long-range interaction present, i.~e.~when either the gauge field acquires
a mass due to a Chern-Simons term or abnormal screening at a high
doping level takes place \cite{wen2}.
\par
For the bound states (\ref{boson-boson})-(\ref{fermion-fermion})
we get the following picture.
At a doping level which is greater than the critical one, there is a
crossover from the QC phase at high temperatures to the QD phase at low $T$.
Below the crossover temperature $T^*$ the bound states will form and this
should be reflected by various physical quantities.
\par
The fact that the formation of the bound states brings the system
towards a singlet state, may explain the anomalous data in nuclear magnetic
relaxation experiments \cite{takig}, where a strong reduction of both the
Knight-shift and the relaxation-rate on oxygen and copper is observed as a
function of temperature, well above the superconducting transition, for
underdoped samples of YBCO. Closely connected with this
is the observation of a spin-gap in neutron scattering experiments on
YBCO \cite{barz} which corresponds to the mass of the $Z$-bosons.
Moreover, when the bound states between a fermion and a boson form,
charge-spin separation takes
place and the scattering of the charge-carriers by the gauge fields vanishes
since the new bound state has zero charge and therefore does not couple
to the gauge field.
Such a scenario is consistent with recent resistivity measurements
in underdoped YBCO 123 \cite{buch,takenaka}
and 124 \cite{itox}, where a reduction
of the resistivity with respect to the linear temperature dependence is
observed, at a temperature $T^*$ where the nuclear relaxation rate
($1/T_1T$) on Cu shows a maximum that is commonly associated with the
spin-gap \cite{takig}. Furthermore in the resistivity perpendicular to the
CuO-planes a crossover from a metallic-like to a semiconductor-like
behavior is observed at the same characteristic temperature $T^*$
\cite{takenaka}, which
in our theory follows from the fact that one has to break the
bound states in order to have charge transport perpendicular to the planes.
This is also consistent with a suppression of the
optical conductivity parallel to the
$c$-axis that was reported recently for underdoped samples of 123 \cite{home}
indicating the opening of a pseudo gap along the $c$-axis.
\par
In conclusion we have shown that within the long-wavelength,
low-energy sector of the spin-fermion model a quantitative correct description
for the doping induced quantum phase transition appearing
in the HTSC materials can be obtained.
Furthermore, in the gauge field theory of the
doped AF in the rotating reference frame, charge-spin separation and pairing
follow in a natural way from the formation of
bound state in the QD phase (above the critical
doping and at low temperatures) due to the long-range force mediated by the
singular mode of the transverse gauge field, such that an intimate
connection between these phenomena and the spin-gap results.
Finally, we have also discussed the
connection of these bound states with the anomalous features observed
in the low-doping regime of the cuprates.
\\[10mm]
We acknowledge support by the Deutsche Forschungsgemeinschaft under Project
No.~Mu 820/5-2.

\begin{appendix}

\section{tensorial decomposition}
\label{tensorial decomposition}
\begin{enumerate}
\renewcommand{\labelenumi}{(\alph{enumi})}
\item {\bf Evaluation of $\Gamma_1^F$}\\
      Due to the properties (\ref{properties}) of the two
      transversal tensors $A_{\mu\nu}$ and $B_{\mu\nu}$ we get
      ($p^2 = \vec p^2 + \omega_n^2$)
      \begin{eqnarray}
      \left( \delta_{\mu 0} - \frac{p_\mu p_0}{p^2} \right)
      \frac{p^2}{\vec p^2}
      \left( \delta_{0 \nu} - \frac{p_0 p_\nu}{p^2} \right)
      \Pi_{\mu\nu}^F (p) = \frac{p^2}{\vec p^2}\Pi_{00}^F(p)=\Gamma_1^F(p)\; ,
      \end{eqnarray}
      where $\Pi_{00}(p)$ follows from the expression (\ref{fermions}) of the
      fermionic contribution to the polarization tensor:
      \begin{eqnarray}
      \Pi_{00}(p)&=&
 \frac{1}{\beta}\sum_{\nu_m}\int \frac{d^2q}{(2\pi)^2}
                   \frac{1}{\left[\frac{1}{2m}\vec q^{\:2} - i\nu_m\right]
                            \left[\frac{1}{2m}\left(\vec q + \vec p\right)^2
                                   -i (\nu_m +\omega_n)\right]}
       =: I_2^F (p) \; .
      \end{eqnarray}
\item {\bf Evaluation of $\Gamma_2^F$}\\
      From (\ref{properties}) further follows:
      \begin{eqnarray}
      \delta_{\mu i}
             \left( \delta_{ij} - \frac{p_i p_j}{\vec p^2} \right)
      \delta_{j \nu}\; \Pi_{\mu\nu}^F =
      \left( \delta_{ij} - \frac{p_i p_j}{\vec p^2} \right)\Pi_{ij}^F
      = \Pi_{ii}^F -\frac{p_ip_j}{\vec p^{\: 2}}\Pi_{ij}^F =\Gamma_2^F\; ,
      \end{eqnarray}
      with
      \begin{eqnarray}
    \Pi_{ii}^F &=& \frac{\delta_{ii}}{m}
                 \frac{1}{\beta}\sum_{\nu_m}\int\frac{d^2q}{(2\pi)^2}
                 \frac{1}{\frac{1}{2m}\vec q^{\: 2}-i\nu_m}
               - \left(\frac{1}{2m}\right)^2
                 \frac{1}{\beta}\sum_{\nu_m}\int\frac{d^2q}{(2\pi)^2}
                 \frac{(2\vec q+\vec p)\cdot (2\vec q+\vec p)}{
                 \left[\frac{1}{2m}\vec q^{\: 2}-i\nu_m\right]
            \left[\frac{1}{2m}\left(\vec q+\vec p\right)^{\: 2}
                  -i \left(\nu_m+\omega_n\right)\right]}
                  \nonumber \\
               &=:& \frac{\delta_{ii}}{m} I_1^F(p)
                - \left(\frac{1}{2m}\right)^2 I_{3}^F(p)
      \end{eqnarray}
      and
      \begin{eqnarray}
      \hspace*{-07mm}
      p_i p_j\Pi_{ij}^F&=&\frac{1}{m} \vec p^2
                 \frac{1}{\beta}\sum_{\nu_m}\int\frac{d^2q}{(2\pi)^2}
                 \frac{1}{\frac{1}{2m}\vec q^{\: 2}-i\nu_m}
                 \nonumber \\
      &-&\left(\frac{1}{2m}\right)^2
      \frac{1}{\beta}\sum_{\nu_m}\int\frac{d^2q}{(2\pi)^2}
      \frac{\left(2\vec q + \vec p\right)\cdot \vec p \hspace*{05mm}
            \left(2\vec q + \vec p\right)\cdot \vec p}
           {\left[\frac{1}{2m}\vec q^{\: 2}-i\nu_m\right]
            \left[\frac{1}{2m}\left(\vec q+\vec p\right)^{\: 2}-i
                              \left(\nu_m+\omega_n\right)\right]}
      \nonumber \\
      &=:& \frac{1}{m}\vec p^2 I_1^F(p)-\left(\frac{1}{2m}\right)^2 \: I_4^F(p)
      \end{eqnarray}
\end{enumerate}

\section{Fermionic and bosonic integrals}
\label{integrals}
In this appendix we calculate the bosonic and fermionic integrals
appearing in the scalar functions $\Gamma^B$, $\Gamma^F_1$ and $\Gamma_2^F$
of the polarization tensor for the gauge field.
\par
First we evaluate the bosonic integral $F(\vec p,\omega_n)$ defined by:
\begin{eqnarray}
F(\vec p,\omega_n)=
\frac{1}{\beta} \sum_{\tilde\nu}\int\frac{d^2q}{(2\pi)^2}
\frac{1}{\left[\vec q^2+\nu_m^2+M^2\right]
\left[(\vec q+\vec p)^2+(\nu_m +\omega_n)^2+M^2\right]} \; .
\end{eqnarray}
Using a Feynman parametrisation we obtain:
\begin{eqnarray}
F(\vec p,\omega_n) =
\frac{1}{\beta} \sum_{\tilde\nu}\int\frac{d^2q}{(2\pi)^2}
\int_0^1 dx\;
\frac{1}{\Bigl[M^2+\vec q^2+\tilde\nu^2+\vec p^2x(1-x)+\omega^2x(1-x)\Bigr]^2}
\end{eqnarray}
where $\tilde\nu=\nu +\omega (1-x)$.
Integration over momentum leads to:
\begin{eqnarray}
F(\vec p,\omega_n)=
\frac{1}{\beta} \sum_{\tilde \nu}
\frac{1}{4\pi}\int_0^1 dx\; \frac{1}{M^2+(\vec
p^2+\omega^2)x(1-x)+\tilde \nu^2}\; .
\end{eqnarray}
In order to evaluate the frequency sum we make a partial fraction
decomposition of the integrand. (Here we have to introduce a factor
$\exp(i\nu\eta)$ to ensure convergence of the series):
\begin{eqnarray}
F(\vec p,\omega_n)&=&
\lim_{\eta\rightarrow 0}
\int_0^1 \frac{dx}{4\pi}\;
\frac{1}{\beta} \sum_{\tilde \nu}
\frac{1}{2\sqrt{M^2+(\vec p^2+\omega_n^2)x(1-x)}}
\exp(i\nu\eta)\cdot \nonumber \\
& &\left\{
\frac{1}{i\nu+\left(\sqrt{M^2+(\vec p^2+\omega_n^2)x(1-x)}
+i\omega_n(1-x)\right)}
-\frac{1}{i\nu-\left(\sqrt{M^2+(\vec p^2+\omega_n^2)x(1-x)}
-i\omega_n(1-x)\right)}
\right\} \nonumber \\
&=& \int_0^1 \frac{dx}{4\pi}\;
\frac{1}{2\sqrt{M^2+(\vec p^2+\omega_n^2)x(1-x)}}\\
& &
\!\!\!\!\!\!\!\!\!\!\!
\!\!\!\!\!\!\!\!\!\!\!
\!\!\!\!\!\!\!\!\!\!\!
\left\{1+
\frac{1}{\exp
\left[\beta\sqrt{M^2+(\vec p^2+\omega_n^2)x(1-x)}
-i\beta\omega_n(1-x)\right]-1} +
\frac{1}{\exp
\left[\beta\sqrt{M^2+(\vec p^2+\omega_n^2)x(1-x)}
+i\beta\omega_n(1-x)\right]-1}
\right\} \nonumber
\end{eqnarray}
For $T\rightarrow 0$
the temperature dependent part of the integral gives only an
exponentially small correction and can therefore be neglected.
After an analytical
continuation to reel frequencies $i\omega_n\rightarrow \omega + i\eta $
we get:
\begin{eqnarray}
F(\vec p,\omega)&\stackrel{T\rightarrow 0}{\simeq}&
\frac{1}{4\pi}\frac{1}{\sqrt{\vec p^2-\omega^2}}
\arcsin\left(\sqrt{\frac{\vec p^2-\omega^2}{4M^2+(\vec p^2-\omega^2)}}\right)
\; .
\end{eqnarray}
\par
In the following we calculate the fermionic integrals.
The first one is gives and equals the density of fermions at the
fermi surface:
\begin{eqnarray}
I_1^F = \frac{1}{\beta}\sum_{\nu_m}\int\frac{d^2q}{(2\pi)^2}\:
         \frac{1}{\frac{1}{2m}\vec q^{\: 2}-i\nu_m}
      =  - \rho_F \; .
\end{eqnarray}
For the calculation of the remaining fermionic integrals we make use
of the formal identity
\begin{eqnarray}
\frac{1}{\omega \pm i \eta} =
{\cal P} \frac{1}{\omega} \mp i \pi \; \delta(\omega)
\end{eqnarray}
with ${\cal P}$ denoting the principal value.
In this way we obtain
\begin{eqnarray}
I_2^F &=& \frac{1}{\beta}\sum_{\nu_m}\int\frac{d^2q}{(2\pi)^2}\:
          \frac{1}
               {\left[\frac{1}{2m}\vec q^{\: 2}-i\nu_m\right]
                \left[\frac{1}{2m}\left(\vec q+\vec p\right)^{\: 2}
                -i \nu_m-\left(\omega+i\eta\right)\right]
               } \nonumber \\
      &=& -{\cal P} \int\frac{d^2q}{(2\pi)^2}\:
          \frac{n(\vec q - \vec p/2) - n(\vec q + \vec p/2)}
               {\omega - \left(\vec q \cdot \vec p\right)/m}
       +  i\pi \int\:\frac{d^2q}{(2\pi)^2}\:
          \left[n(\vec q - \vec p/2) - n(\vec q + \vec p/2)\right]
          \delta\left(\omega-\left(\vec q \cdot \vec p\right)/m\right)
\; .
\end{eqnarray}
Furthermore, we go over to zero temperature, where the difference
between the two Fermi distributions is given by
\begin{eqnarray}
n(\vec q - \vec p/2) - n(\vec q + \vec p/2)= -\vec p \cdot \vec \nabla
n(\vec q) \stackrel{T=0}{=} \vec p \cdot \vec q/q \:
\delta (\vert \vec q\vert - k_F) \; .
\end{eqnarray}
Using this expression we get
\begin{eqnarray}
I_2^F &=& -{\cal P} (\frac{1}{2\pi})^2\:
          2 \int_0^\pi d\theta \int_0^\infty dq \:
          \delta(q -k_F)
          \frac{qp \cos \theta}{\omega-\left(q p \cos \theta\right)/m}
          \nonumber \\
      &+& i\frac{1}{2\pi} \int_0^\pi d\theta \int_0^\infty dq \:
          qp \cos\theta\: \delta(q -k_F)
          \delta\left(\omega-\left(qp\cos \theta\right)/m\right)
          \nonumber \\
      &=& 2m \left(\frac{1}{2\pi}\right)^2 {\cal P} \int_{-1}^1 dz \:
          \frac{z}{z-x}\;
          \frac{1}{\sqrt{1-z^2}}
       +i (\frac{1}{2\pi}) m \int_{-1}^1 dz \:
          \delta \left(z-x\right)
          \frac{z}{\sqrt{1-z^2}}
          \nonumber \\
      &=& \frac{m}{2\pi}
          \left( 1 - \frac{x\Theta(x^2-1)} {\sqrt{x^2-1}}
                 +i \frac{x\Theta(1-x^2)} {\sqrt{1-x^2}}
          \right) \; \; ; \; \; x=\left|\frac{\omega}{\vec p v_F}\right|\; .
\end{eqnarray}
The remaining fermionic integrals are calculated in the same way and we
just give the results here
\begin{eqnarray}
I_3^F &=& \frac{1}{\beta}\sum_{\nu_m}\int\frac{d^2q}{(2\pi)^2}\:
          \frac{(2\vec q +\vec p)\cdot (2\vec q + \vec p)}
               {\left[\frac{1}{2m}\vec q^{\: 2}-i\nu_m\right]
                \left[\frac{1}{2m}\left(\vec q+\vec p\right)^{\: 2}
                -i \nu_m-\left(\omega+i\eta\right)\right] }
          \nonumber \\
      &=& 4 k_F^2 I_2^F
\end{eqnarray}
and
\begin{eqnarray}
I^F_4 &=&
\frac{1}{\beta}\sum_{\nu_m}\int\frac{d^2q}{(2\pi)^2}
      \frac{\left(2\vec q + \vec p\right)\cdot \vec p \hspace*{05mm}
            \left(2\vec q + \vec p\right)\cdot \vec p}
           {\left[\frac{1}{2m}\vec q^{\: 2}-i\nu_m\right]
            \left[\frac{1}{2m}\left(\vec q+\vec p\right)^{\: 2}-i
                              \nu_m-\left(\omega+i\eta\right)\right]}
\nonumber \\
&=& \frac{m}{2\pi} 4k_F^2 \vec p^2
    \left(
         x^2 + \frac{1}{2} - \frac{x^3\Theta(x^2-1)} {\sqrt{x^2-1}}
         +i \frac{x^3\Theta(1-x^2)} {\sqrt{1-x^2}}
    \right) \; .
\end{eqnarray}

\end{appendix}

\end{document}